\documentclass{achemso}
\setkeys{acs}{articletitle = true, maxauthors=0}
\SectionNumbersOn

\usepackage{chemformula} % Formula subscripts using \ch{}
\usepackage[T1]{fontenc} % Use modern font encodings
\usepackage{multicol}
\usepackage{booktabs}
\usepackage{rotating}
\usepackage{amsmath}\allowdisplaybreaks[1]
\usepackage{braket}
\usepackage{xcolor}
\usepackage{mathrsfs}
\usepackage{simpler-wick}
\usepackage{cancel}
\usepackage{siunitx}

\def\hhoo{(\textit{P})-hydrogen peroxide}
\def\smo{(\textit{S})-methyloxirane}

\author{Brendan M. Shumberger}
\author{Kirk C. Pearce}
\author{T. Daniel Crawford}
\email{crawdad@vt.edu}
\affiliation{Department of Chemistry, Virginia Tech, Blacksburg, Virginia, U.S.A.}

\title{Analytic Computation of Vibrational Circular Dichroism Spectra Using Second-Order M{\o}ller-Plesset Perturbation Theory}

\abbreviations{AATs, MP2}

\begin{document}

\begin{abstract}
We present the first analytic-derivative-based formulation of vibrational circular dichroism (VCD) atomic axial tensors for
second-order M{\o}ller-Plesset (MP2) perturbation theory.  We compare our implementation to our recently reported
finite-difference approach and find close agreement, thus validating the new formulation. The new approach is dramatically
less computationally expensive than the numerical-derivative method with an overall computational scaling of ${\cal
O}(N^6)$.  In addition, we report the first fully analytic VCD spectrum for \smo\ at the MP2 level of theory.
\end{abstract}

\section{Introduction}

From its advent, vibrational circular dichroism (VCD) --- the differential absorption of left- and
right-circularly-polarized infrared light by a chiral compound --- has been a challenge to quantum chemistry because the
requisite rotatory strengths vanish within the Born-Oppenheimer approximation.  The VCD rotatory strength is obtained from
the dot product of the electric- and magnetic-dipole transition-moment vectors between vibrational states, and while the
former may be straightforwardly computed via differentation of the expectation value of the electric dipole moment operator
in the electronic ground state, the same approach fails for the latter because the corresponding expectation value is zero
for closed-shell states.  In the 1970s and 1980s, a number of \textit{ad hoc} models were put forward in an attempt to
simulate VCD spectra \cite{Holzwarth1972, Schellman1973, Nafie1977, Abbate1981, Nafie1983a, Nafie1986, Freedman1983,
Polavarapu1983, Barnett1980, Barron1979}, but it was Stephens's 1985 formulation\cite{Stephens1985a} of the magnetic-dipole
transition moment [commonly referred to as the atomic axial tensor (AAT)] that provided the first, general ground-state VCD
approach, requiring only the overlap of wave function derivatives with respect to nuclear coordinates and the external
magnetic-field.

The first computations of VCD AATs using Stephens's approach were reported at the Hartree-Fock (HF) level by Lowe, Segal,
and Stephens in 1986 using numerical differentiation of the ground-state wave function\cite{Lowe1986, Lowe86:CPL}.  While
this approach provided useful insights and benchmarks, it was not sufficient for practical calculations due to the need for
complex wave function representations (and algebra) for finite magnetic-field perturbations.  In 1987, Amos, Handy,
Jalkanen, and Stephens\cite{Amos1987} described the first implementation of analytic-derivative techniques for the
calculation of HF-level AATs, requiring solution of the first-order coupled-perturbed Hartree-Fock (CPHF) equations for the
derivatives of the molecular orbital (MO) coefficients, as well as half-derivative overlap integrals.  Six years later, Bak
\textit{et al.}\ reported their extension of Stephens's AAT formulation to the multiconfigurational self-consistent field
(MCSCF) level of theory \cite{Bak1993} allowing for the inclusion of static electron correlation effects while
simultaneously tackling the origin independence problem by introducing gauge-including atomic orbitals
(GIAOs).\cite{Ditchfield1971, Helgaker91} Soon thereafter, Stephens and co-workers\cite{Stephens94:B3LYP} simulated VCD
spectra using second-order M{\o}ller-Plesset (MP2) theory and density functional theory (DFT) though only the harmonic force
fields and the electronic-dipole transition moments [atomic polar tensors (APT)] were computed using these methods; the AAT
was computed only at the HF level of theory.  Building upon this work, in 1996 Cheeseman \textit{et al.}\cite{Cheeseman1996}
carried out the first full simulation of VCD spectra using DFT, including the use of GIAOs.

Recently, we reported the first simulations of VCD spectra at the MP2 and configuration interaction doubles (CID) levels of
theory.\cite{Shumberger2024}  Similar to Stephens's approach in the mid-1980s using HF theory, we carried out the
calculation of the AATs using numerical differentiation of the corresponding wave functions.  However, the required
determinantal expansions lead to much greater complexity and computational cost in the MP2/CID cases compared to HF due to
the need to evaluate overlaps between doubly excited determinants in non-orthonormal bra and ket bases.  This yields an
${\cal O}(N^{11})$ scaling of the method, thus limiting our analysis to only small molecules with modest basis sets.

Here, we present the first analytic-derivative approach to computing MP2 AATs.  Our implementation eliminates the need for
complex arithmetic and non-orthonormal MOs, yielding an overall scaling of ${\cal O}(N^6)$.  In the next section, we will
outline our derivation of the working equations, followed by validation of the results by comparison between
finite-difference and analytic AATs.  We then report the first MP2 VCD spectra simulations for (\textit{S})-methyloxirane
for several basis sets.

\section{Theory}

In Stephens's formulation of VCD rotatory strengths,\cite{Stephens1985a} the electronic contribution to the AAT,
$I_{\alpha\beta}^{\lambda}$, is obtained from the overlap between derivatives of the ground state wave function, $\Psi_G$,
with respect to nuclear displacement, $R_{\lambda\alpha}$, and the external magnetic field, $H_{\beta}$,
\begin{equation} \label{ElectronicAAT}
    I_{\alpha\beta}^{\lambda} = \left\langle
    \left( \frac{\partial \Psi_G(\vec{R})}{\partial R_{\lambda\alpha}} \right)_{R_{\lambda\alpha} = R_{\lambda\alpha}^0} \bigg|
    \left( \frac{\partial \Psi_G(\vec{R}_0, H_{\beta})}{\partial H_{\beta}} \right)_{H_{\beta}=0} \right\rangle,
\end{equation}
which is evaluated at the equilibrium geometry and at zero field.
The indices $\lambda$ and $\alpha$ denote the nucleus and
Cartesian axis of the $\lambda^{th}$ nucleus' displacement, respectively, while
$\beta$ denotes the Cartesian axis of the magnetic field. By approximating the
wave function using first-order M{\o}ller-Plesset theory,
\begin{equation} \label{MP2WaveFunction}
  \left| \Psi_G(R, H_{\beta}) \right\rangle \approx \left( 1 + \hat{T}_2 \right) \left| \Phi_0 \right\rangle,
\end{equation}
we can obtain the second-order M{\o}ller-Plesset (MP2) AAT where the
$\hat{T}_2$ operator is
\begin{equation} \label{Amplitudes}
    \hat{T}_2 = \frac{1}{4} \sum_{ijab} t_{ij}^{ab} a_a^{\dagger} a_b^{\dagger} a_j a_i.
\end{equation}
and $\Phi_0$ is the HF reference determinant. We use the standard indexing
scheme where $i$, $j$, $k$, \ldots denote occupied spin-orbitals, and $a$, $b$, $c$, \ldots denote virtual spin-orbitals.
Indices $p$ and $q$ will denote MOs appearing in either subspace. Inserting 
Eqs.~\eqref{MP2WaveFunction} -~\eqref{Amplitudes} into Eq.~\eqref{ElectronicAAT}
with subsequent application of the second quantized operators onto the HF
reference determinant yields the intermediately normalized AAT (denoted by the
subscript $int$) of the form
\begin{align} \label{MP2AATs}
    \left[ I_{\alpha\beta}^{\lambda} \right]_{int} & = 
    \left\langle \frac{\partial \Phi_{0}}{\partial R_{\lambda\alpha}} \bigg| \frac{\partial \Phi_{0}}{\partial H_{\beta}} \right\rangle
    + \frac{1}{4} \sum_{ijab} \frac{\partial {t_{ij}^{ab}}^\dagger}{\partial R_{\lambda\alpha}}
    \left\langle \Phi_{ij}^{ab} \bigg| \frac{\partial \Phi_{0}}{\partial H_{\beta}} \right\rangle
    + \frac{1}{4} \sum_{ijab} {t_{ij}^{ab}}^\dagger
    \left\langle \frac{\partial \Phi_{ij}^{ab}}{\partial R_{\lambda\alpha}} \bigg| \frac{\partial \Phi_{0}}{\partial H_{\beta}} \right\rangle \nonumber \\
    & + \frac{1}{4} \sum_{ijab} \frac{\partial t_{ij}^{ab}}{\partial H_{\beta}}
    \left\langle \frac{\partial \Phi_{0}}{\partial R_{\lambda\alpha}} \bigg| \Phi_{ij}^{ab} \right\rangle
    + \frac{1}{4} \sum_{ijab} t_{ij}^{ab} 
    \left\langle \frac{\partial \Phi_{0}}{\partial R_{\lambda\alpha}} \bigg| \frac{\partial \Phi_{ij}^{ab}}{\partial H_{\beta}} \right\rangle \nonumber \\
    & + \frac{1}{16} \sum_{ijab} \sum_{klcd}
    \frac{\partial t{_{ij}^{ab}}^\dagger}{\partial R_{\lambda\alpha}} \frac{\partial t_{kl}^{cd}}{\partial H_{\beta}}
    \left\langle \Phi_{ij}^{ab} \bigg| \Phi_{kl}^{cd} \right\rangle
    + \frac{1}{16} \sum_{ijab} \sum_{klcd} \frac{\partial {t_{ij}^{ab}}^\dagger}{\partial R_{\lambda\alpha}} t_{kl}^{cd}
    \left\langle \Phi_{ij}^{ab} \bigg| \frac{\partial \Phi_{kl}^{cd}}{\partial H_{\beta}} \right\rangle \nonumber \\
    & + \frac{1}{16} \sum_{ijab} \sum_{klcd} {t_{ij}^{ab}}^\dagger \frac{\partial t_{kl}^{cd}}{\partial H_{\beta}}
    \left\langle \frac{\partial \Phi_{ij}^{ab}}{\partial R_{\lambda\alpha}} \bigg| \Phi_{kl}^{cd} \right\rangle
    + \frac{1}{16} \sum_{ijab} \sum_{klcd} {t_{ij}^{ab}}^\dagger t_{kl}^{cd}
    \left\langle \frac{\partial \Phi_{ij}^{ab}}{\partial R_{\lambda\alpha}} \bigg| \frac{\partial \Phi_{kl}^{cd}}{\partial H_{\beta}} \right\rangle
\end{align}
where $\Phi_{ij}^{ab}$ is a doubly substituted determinant. Evaluation of the overlap between wave function derivatives can
be performed by considering the derivative of an MO, $\varphi_p$,
\begin{equation}
    \frac{\partial \varphi_p}{\partial \chi} = \sum_i U_{ip}^{\chi} \varphi_i + \sum_a U_{ap}^{\chi} \varphi_a + \varphi_p^{\chi}. \label{MODerivative}
\end{equation}
In Eq.~\eqref{MODerivative}, $U_{qp}^{\chi}$ is a CPHF coefficient and $\varphi_p^{\chi}$ is the derivative of an atomic
orbital (AO) with respect to some arbitrary perturbation $\chi$, transformed into the MO basis (also known as a core
derivative). Solutions to the CPHF equations for nuclear displacements and magnetic field perturbations have been reviewed
by Yamaguchi et al.\cite{Yamaguchi1994} and by Amos et al.\cite{Amos1987}, among others, and we direct readers to these
works for further information. Utilizing the derivative product rule on the Slater determinant,
\begin{equation} \label{DeterminantExpansion}
    \left| \frac{\partial \Phi_0}{\partial \chi} \right\rangle
    = \left| \frac{\partial \varphi_1}{\partial \chi} \varphi_2 ... \varphi_N \right\rangle
    + \left| \varphi_1 \frac{\partial \varphi_2}{\partial \chi} ... \varphi_N \right\rangle + ...
    + \left| \varphi_1 \varphi_2 ... \frac{\partial \varphi_N}{\partial \chi} \right\rangle,
\end{equation}
and inserting Eq.~\eqref{MODerivative}, we can define the derivative of our
reference wave function in a second-quantized notation
\begin{align}
    \left| \frac{\partial \Phi_0}{\partial \chi} \right\rangle
    &= \sum_k U_{k1}^{\chi} a_k^{\dagger} a_1^{} \left| \Phi_0 \right\rangle
    + \sum_c U_{c1}^{\chi} a_c^{\dagger} a_1^{} \left| \Phi_0 \right\rangle
    + a_{1_{\chi}}^{\dagger} a_1^{} \left| \Phi_0 \right\rangle \nonumber \\
    &+ \sum_k U_{k2}^{\chi} a_k^{\dagger} a_2 \left| \Phi_0 \right\rangle
    + \sum_c U_{c2}^{\chi} a_c^{\dagger} a_2 \left| \Phi_0 \right\rangle
    + a_{2_{\chi}}^{\dagger} a_2 \left| \Phi_0 \right\rangle + ... \nonumber \\
    &+ \sum_k U_{kN}^{\chi} a_k^{\dagger} a_N \left| \Phi_0 \right\rangle
    + \sum_c U_{cN}^{\chi} a_c^{\dagger} a_N \left| \Phi_0 \right\rangle
    + a_{N_{\chi}}^{\dagger} a_N \left| \Phi_0 \right\rangle,
\end{align}
where $a_{p_{\chi}}^{\dagger}$ is a creation operator for the derivative of $\varphi_p$ with respect to $\chi$.  (Note that
the derivatives of the spin-orbitals are not orthogonal to their unperturbed counterparts.) This expression can be reduced
to
\begin{align}
    \left| \frac{\partial \Phi_0}{\partial \chi} \right\rangle
    &= \sum_i U_{ii}^{\chi} \left| \Phi_0 \right\rangle
    + \sum_i \sum_c U_{ci}^{\chi} a_c^{\dagger} a_i^{} \left| \Phi_0 \right\rangle
    + \sum_i a_{i_{\chi}}^{\dagger} a_i^{} \left| \Phi_0 \right\rangle. \label{GroundSDDerivative}
\end{align}
Similar to the derivative of the reference determinant, we can express the
derivative of the doubly excited determinant in a second quantized notation as,
\begin{align}
    \left| \frac{\partial \Phi_{ij}^{ab}}{\partial \chi} \right\rangle
    & = \left( \sum_{l \ne i,j} U_{ll}^{\chi} + U_{aa}^{\chi} + U_{bb}^{\chi} \right)  \left| \Phi_{ij}^{ab} \right\rangle
     - \sum_{l \ne i,j} \left( U_{il}^{\chi} \left| \Phi_{lj}^{ab} \right\rangle + U_{jl}^{\chi} \left| \Phi_{il}^{ab} \right\rangle \right) \nonumber \\
    &+ \left( U_{ia}^{\chi} \left| \Phi_j^b \right\rangle - U_{ja}^{\chi} \left| \Phi_i^b \right\rangle
     - U_{ib}^{\chi} \left| \Phi_j^a \right\rangle + U_{jb}^{\chi} \left| \Phi_i^a \right\rangle \right) \nonumber \\
    &+ \sum_{l \ne i,j} \sum_{c \ne a,b} U_{cl}^{\chi} \left| \Phi_{ijl}^{abc} \right\rangle
     + \sum_{c \ne a,b} \left( U_{ca}^{\chi} \left| \Phi_{ij}^{cb} \right\rangle + U_{cb}^{\chi} \left| \Phi_{ij}^{ac} \right\rangle \right) \nonumber \\
    &+ \sum_{l \ne i,j} a_{l_{\chi}}^{\dagger} a_l^{} \left| \Phi_{ij}^{ab} \right\rangle + a_{a_{\chi}}^{\dagger} a_a^{} \left| \Phi_{ij}^{ab} \right\rangle + a_{b_{\chi}}^{\dagger} a_b^{} \left| \Phi_{ij}^{ab} \right\rangle, \label{DoublesSDDerivative}
\end{align}
where we have separated operator strings by their action on the reference determinant and removed redundant operator pairs.
With our derivative determinants in Eqs.~\eqref{GroundSDDerivative} and ~\eqref{DoublesSDDerivative}, we can write the
explicit forms for the bra-state derivatives with respect to nuclear displacements and ket-states derivatives with respect
to magnetic-field perturbations. For the HF reference determinant, we obtain
\begin{align}
    \left\langle \frac{\partial \Phi_0}{\partial R_{\lambda\alpha}} \right|
    &= \sum_m U_{mm}^{R_{\lambda\alpha}} \left\langle \Phi_0 \right|
    + \sum_m \sum_e U_{em}^{R_{\lambda\alpha}} \left\langle \Phi_0 \right| \{ a_m^{\dagger} a_e^{} \}
    + \sum_m \left\langle \Phi_0 \right| a_m^{\dagger} a_{m_{R_{\lambda\alpha}}}^{}, \label{NucRef}
\end{align}
and
\begin{align}
    \left| \frac{\partial \Phi_0}{\partial H_{\beta}} \right\rangle
    &= \sum_n U_{nn}^{H_\beta} \left| \Phi_0 \right\rangle
    + \sum_n \sum_f U_{fn}^{H_\beta} \{ a_f^{\dagger} a_n^{} \} \left| \Phi_0 \right\rangle. \label{MagRef}
\end{align}
Note that terms involving core derivatives do not appear in Eq.~\eqref{MagRef} because we are not including field-dependent
basis functions (GIAOs) in the current formulation.  For the corresponding derivatives of the doubly excited determinants,
including the wave function amplitudes and associated spin-orbital summations allows us to take advantage of the symmetry of
the indices yielding the simplified expression,
\begin{align} \label{NucT2_1}
    \frac{1}{4} & \sum_{ijab} {t_{ij}^{ab}}^\dagger \left\langle \frac{\partial \Phi_{ij}^{ab}}{\partial R_{\lambda\alpha}} \right|
     = \frac{1}{4} \sum_{ijab} {t_{ij}^{ab}}^\dagger \Bigg[
       \left( \sum_{m \ne i,j} U_{mm}^{R_{\lambda\alpha}} + 2 U_{aa}^{R_{\lambda\alpha}} \right) \left\langle \Phi_0 \right| \{ a_i^{\dagger} a_j^{\dagger} a_b^{} a_a^{} \} \nonumber \\
    &- 2 \sum_{m \ne i,j} U_{im}^{R_{\lambda\alpha}} \left\langle \Phi_0 \right| \{ a_m^{\dagger} a_j^{\dagger} a_b^{} a_a^{} \}
     + 2 \sum_{e \ne a,b} U_{ea}^{R_{\lambda\alpha}} \left\langle \Phi_0 \right| \{ a_i^{\dagger} a_j^{\dagger} a_b^{} a_e^{} \} \nonumber \\
    &+ \sum_{m \ne i,j} \sum_{e \ne a,b} U_{em}^{R_{\lambda\alpha}} \left\langle \Phi_0 \right| \{ a_i^{\dagger} a_j^{\dagger} a_m^{\dagger} a_e^{} a_b^{} a_a^{} \}
     + 4 U_{ia}^{R_{\lambda\alpha}} \left\langle \Phi_0 \right| \{ a_j^{\dagger} a_b^{} \} \nonumber \\
    &  + \sum_{m \ne i,j}  \left\langle \Phi_0 \right| \{ a_i^{\dagger} a_j^{\dagger} a_b^{} a_a^{} \} a_m^{\dagger} a_{m_{R_{\lambda\alpha}}}^{}
     + 2 \left\langle \Phi_0 \right| \{ a_i^{\dagger} a_j^{\dagger} a_b^{} a_a^{} \} a_a^{\dagger} a_{a_{R_{\lambda\alpha}}}^{}
    \Bigg]
\end{align}
and
\begin{align} \label{MagT2_1}
    \frac{1}{4} & \sum_{klcd} t_{kl}^{cd} \left| \frac{\partial \Phi_{kl}^{cd}}{\partial H_{\beta}} \right\rangle
     = \frac{1}{4} \sum_{klcd} t_{kl}^{cd} \Bigg[
       \left( \sum_{n \ne k,l} U_{nn}^{H_{\beta}} + 2 U_{cc}^{H_{\beta}} \right) \{ a_c^{\dagger} a_d^{\dagger} a_l^{} a_k^{} \} \left| \Phi_0 \right\rangle \nonumber \\
    &- 2 \sum_{n \ne k,l} U_{kn}^{H_{\beta}} \{ a_c^{\dagger} a_d^{\dagger} a_l^{} a_n^{} \} \left| \Phi_0 \right\rangle
     + 2 \sum_{f \ne c,d} U_{fc}^{H_{\beta}} \{ a_f^{\dagger} a_d^{\dagger} a_l^{} a_k^{} \} \left| \Phi_0 \right\rangle \nonumber \\
    &+ \sum_{n \ne k,l} \sum_{f \ne c,d} U_{fn}^{H_{\beta}} \{ a_c^{\dagger} a_d^{\dagger} a_f^{\dagger} a_n^{} a_l^{} a_k^{} \} \left| \Phi_0 \right\rangle
     + 4 U_{kc}^{H_{\beta}} \{ a_d^{\dagger} a_l^{} \} \left| \Phi_0 \right\rangle
    \Bigg].
\end{align}
The overlaps between Eqs.~\eqref{NucRef}, \eqref{MagRef}, \eqref{NucT2_1}, and \eqref{MagT2_1} may be evaluated with the
standard Wick's theorem contraction rules\cite{Crawford00:review}, as well as additional rules for the core derivatives,
\textit{viz.},
\begin{align}
    \wick{ \c1 a_i^{\dagger} \c1 a_{q_{R_{\lambda\alpha}}}^{} } &= a_i^{\dagger} a_{q_{R_{\lambda\alpha}}}^{} - \{ a_i^{\dagger} a_{q_{R_{\lambda\alpha}}}^{} \} = a_i^{\dagger} a_{q_{R_{\lambda\alpha}}}^{} + a_{q_{R_{\lambda\alpha}}}^{} a_i^{\dagger} = \left\langle \varphi_q^{{R_{\lambda\alpha}}} | \varphi_i^{} \right\rangle, \\
    \wick{ \c1 a_{q_{R_{\lambda\alpha}}}^{} \c1 a_a^{\dagger} } &= a_{q_{R_{\lambda\alpha}}}^{} a_a^{\dagger} - \{ a_{q_{R_{\lambda\alpha}}}^{} a_a^{\dagger} \} = a_{q_{R_{\lambda\alpha}}}^{} a_a^{\dagger} + a_a^{\dagger} a_{q_{R_{\lambda\alpha}}}^{} = \left\langle \varphi_q^{{R_{\lambda\alpha}}} | \varphi_a^{} \right\rangle, \\
    \wick{ \c1 a_{q_{R_{\lambda\alpha}}}^{\dagger} \c1 a_i^{} } &= a_{q_{R_{\lambda\alpha}}}^{\dagger} a_i^{} - \{ a_{q_{R_{\lambda\alpha}}}^{\dagger} a_i^{} \} = a_{q_{R_{\lambda\alpha}}}^{\dagger} a_i^{} + a_i^{} a_{q_{R_{\lambda\alpha}}}^{\dagger} = \left\langle \varphi_i^{} | \varphi_q^{{R_{\lambda\alpha}}} \right\rangle,
\end{align}
and
\begin{align}
    \wick{ \c1 a_a^{} \c1 a_{q_{R_{\lambda\alpha}}}^{\dagger} } &= a_a^{} a_{q_{R_{\lambda\alpha}}}^{\dagger} - \{ a_a^{} a_{q_{R_{\lambda\alpha}}}^{\dagger} \} = a_a^{} a_{q_{R_{\lambda\alpha}}}^{\dagger} + a_{q_{R_{\lambda\alpha}}}^{\dagger} a_a^{} = \left\langle \varphi_a^{} | \varphi_q^{{R_{\lambda\alpha}}} \right\rangle,
\end{align}
where the terms on the far right-hand side of each equation are half-derivative overlap integrals.
With these equations in hand, we may now derive the various contributions to Eq.~\eqref{MP2AATs} using Wick's theorem and
retaining only the fully contracted terms. The first term is the HF contribution, which evaluates to
\begin{align}
    \left\langle \frac{\partial \Phi_0}{\partial R_{\lambda\alpha}} \bigg| \frac{\partial \Phi_0}{\partial H_{\beta}} \right\rangle
     = \sum_m \sum_e U_{em}^{H_\beta} U_{em}^{R_{\lambda\alpha}}
     + \sum_m \sum_e U_{em}^{H_\beta} \left\langle \varphi_m^{{R_{\lambda\alpha}}} | \varphi_e^{} \right\rangle \label{AATHF}
\end{align}
where we have used the relationship
\begin{align}
    U_{pq}^{R_{\lambda\alpha}} + U_{qp}^{R_{\lambda\alpha}} + S_{pq}^{R_{\lambda\alpha}} = 0,
\end{align}
which is obtained from the derivative of the MO orthonormality condition. The second and fourth terms in Eq.~\eqref{MP2AATs} include
the derivative of the reference determinant projected onto the doubly excited determinant. Since the derivative of the
reference determinant includes at most single excitations [\textit{cf.} Eqs.~\eqref{NucRef} and \eqref{MagRef}], no full
contractions can be produced from these terms. Thus,
\begin{align}
    \frac{1}{4} \sum_{ijab} \frac{\partial {t_{ij}^{ab}}^\dagger}{\partial R_{\lambda\alpha}}
    \left\langle \Phi_{ij}^{ab} \bigg| \frac{\partial \Phi_{0}}{\partial H_{\beta}} \right\rangle
    &= 0 \label{AAT2}
\end{align}
and
\begin{align}
    \frac{1}{4} \sum_{ijab} \frac{\partial t_{ij}^{ab}}{\partial H_{\beta}}
    \left\langle \frac{\partial \Phi_{0}}{\partial R_{\lambda\alpha}} \bigg| \Phi_{ij}^{ab} \right\rangle
    &= 0. \label{AAT4}
\end{align}
We find that the third term,
\begin{align}
    \frac{1}{4} \sum_{ijab} {t_{ij}^{ab}}^\dagger
    \left\langle \frac{\partial \Phi_{ij}^{ab}}{\partial R_{\lambda\alpha}} \bigg| \frac{\partial \Phi_{0}}{\partial H_{\beta}} \right\rangle
    &= \frac{1}{4} \sum_{ijab} {t_{ij}^{ab}}^\dagger \bigg[
       4 U_{bj}^{H_\beta} U_{ia}^{R_{\lambda\alpha}} 
     + 4 U_{bj}^{H_\beta} \left\langle \varphi_a^{{R_{\lambda\alpha}}} | \varphi_i^{} \right\rangle
    \bigg], \label{AAT3}
\end{align}
and fifth term,
\begin{align}
    \frac{1}{4} \sum_{ijab} t_{ij}^{ab}
    \left\langle \frac{\partial \Phi_{0}}{\partial R_{\lambda\alpha}} \bigg| \frac{\partial \Phi_{ij}^{ab}}{\partial H_{\beta}} \right\rangle
    &= - \frac{1}{4} \sum_{ijab} t_{ij}^{ab} \bigg[
       4 U_{jb}^{H_{\beta}} U_{ia}^{R_{\lambda\alpha}}
     + 4 U_{jb}^{H_{\beta}} \left\langle \varphi_a^{{R_{\lambda\alpha}}} | \varphi_i^{} \right\rangle
    \bigg], \label{AAT5}
\end{align}
in Eq.~\eqref{MP2AATs} exactly cancel. As such, only the HF term provides any contribution to the AAT from among the first five
terms. The sixth term includes only derivatives of $\hat{T}$-amplitudes and thus simplifies to
\begin{align}
    \frac{1}{16} \sum_{ijab} \sum_{klcd} & \frac{\partial t{_{ij}^{ab}}^\dagger}{\partial R_{\lambda\alpha}} \frac{\partial t_{kl}^{cd}}{\partial H_{\beta}}
    \left\langle \Phi_{ij}^{ab} \bigg| \Phi_{kl}^{cd} \right\rangle
    = \frac{1}{4} \sum_{ijab} \frac{\partial t{_{ij}^{ab}}^\dagger}{\partial R_{\lambda\alpha}} \frac{\partial t_{ij}^{ab}}{\partial H_{\beta}}. \label{AAT6}
\end{align}
The seventh and eighth terms in Eq.~\eqref{MP2AATs} are
\begin{align}
    \frac{1}{16} \sum_{ijab} \sum_{klcd} & \frac{\partial {t_{ij}^{ab}}^\dagger}{\partial R_{\lambda\alpha}} t_{kl}^{cd}
    \left\langle \Phi_{ij}^{ab} \bigg| \frac{\partial \Phi_{kl}^{cd}}{\partial H_{\beta}} \right\rangle
     = \frac{1}{2} \sum_{ijab} \frac{\partial {t_{ij}^{ab}}^\dagger}{\partial R_{\lambda\alpha}}  \times \nonumber \\
    &  \bigg[ \frac{1}{2} \sum_{n} U_{nn}^{H_{\beta}} t_{ij}^{ab}
     - \sum_{k} U_{ki}^{H_{\beta}} t_{kj}^{ab}
     + \sum_{c} U_{ac}^{H_{\beta}} t_{ij}^{cb} \bigg] \label{AAT7}
\end{align}
and
\begin{align}
    \frac{1}{16} \sum_{ijab} \sum_{klcd} & {t_{kl}^{cd}}^\dagger \frac{\partial t_{ij}^{ab}}{\partial H_{\beta}}
    \left\langle \frac{\partial \Phi_{kl}^{cd}}{\partial R_{\lambda\alpha}} \bigg| \Phi_{ij}^{ab} \right\rangle
     = \frac{1}{2} \sum_{ijab} \frac{\partial t_{ij}^{ab}}{\partial H_{\beta}} \times \nonumber \\
    &  \bigg[
     - \sum_{k} \left( U_{ki}^{R_{\lambda\alpha}}
     + \left\langle \varphi_i^{R_{\lambda\alpha}} | \varphi_k^{} \right\rangle \right) {t_{kj}^{ab}}^\dagger
     + \sum_{c} \left( U_{ac}^{R_{\lambda\alpha}}
     + \left\langle \varphi_c^{R_{\lambda\alpha}} | \varphi_a^{} \right\rangle \right) {t_{ij}^{cb}}^\dagger \bigg], \label{AAT8}
\end{align}
respectively.
The final term in Eq.~\eqref{MP2AATs} involves the overlap between doubly excited derivative determinants and, as a result,
is the most complicated in terms of Wick's theorem contractions. The resulting expression is
\begin{align}
    \frac{1}{16} & \sum_{ijab} \sum_{klcd} {t_{ij}^{ab}}^\dagger t_{kl}^{cd}
    \left\langle \frac{\partial \Phi_{ij}^{ab}}{\partial R_{\lambda\alpha}} \bigg| \frac{\partial \Phi_{kl}^{cd}}{\partial H_{\beta}} \right\rangle
     = \frac{1}{2} \sum_{ijab} {t_{ij}^{ab}}^\dagger \times \nonumber \\
    &  \bigg[
       \sum_{m} \sum_{k} U_{km}^{H_{\beta}} \left( U_{im}^{R_{\lambda\alpha}} + \left\langle \varphi_m^{R_{\lambda\alpha}} | \varphi_i^{} \right\rangle \right) t_{kj}^{ab}
     + \sum_{e} \sum_{c} U_{ec}^{H_{\beta}} \left( U_{ea}^{R_{\lambda\alpha}} + \left\langle \varphi_a^{R_{\lambda\alpha}} | \varphi_e^{} \right\rangle \right) t_{ij}^{cb} \nonumber \\
    &+ \frac{1}{2} \sum_{m} \sum_{e} U_{em}^{H_{\beta}} \left( U_{em}^{R_{\lambda\alpha}} + \left\langle \varphi_m^{R_{\lambda\alpha}} | \varphi_e^{} \right\rangle \right) t_{ij}^{ab}
     - \sum_{m} \sum_{e} U_{ej}^{H_{\beta}} \left( U_{em}^{R_{\lambda\alpha}} + \left\langle \varphi_m^{R_{\lambda\alpha}} | \varphi_e^{} \right\rangle \right) t_{im}^{ab} \nonumber \\
    &- \sum_{m} \sum_{e} U_{bm}^{H_{\beta}} \left( U_{em}^{R_{\lambda\alpha}} + \left\langle \varphi_m^{R_{\lambda\alpha}} | \varphi_e^{} \right\rangle \right)  t_{ij}^{ae}
    \bigg] \label{AAT9}
\end{align}
where we note that $U_{pq}^{H_{\beta}}$ is symmetric and the unperturbed $\hat{T}$-amplitudes are real which leads to the
self-cancellation of several terms involved in the derivation.

The combination of Eqs.~\eqref{AATHF} and \eqref{AAT6}-\eqref{AAT9} results in the analytic spin-orbital expression for the
AAT assuming intermediate normalization of the ground-state wave function. However, Stephens's formulation for the AAT
assumes fully normalized wave functions, which we thus include in our first-order MP wave function as,
\begin{align} \label{MP2WaveFunctionNormalized}
    \left| \Psi_G(R, H_{\beta}) \right\rangle \approx N \left( 1 + \hat{T}_2 \right) \left| \Phi_0 \right\rangle.
\end{align}
Differentiation of this wave function leads to 
\begin{align} \label{ElectronicAATNorm1}
    \left[ I_{\alpha\beta}^{\lambda} \right]_{full}
    &= N^2 \left[ I_{\alpha\beta}^{\lambda} \right]_{int}
     + N \frac{\partial N}{\partial R_{\lambda\alpha}} \bigg[ \left\langle \Phi_0 \bigg| \frac{\partial \Phi_0}{\partial H_{\beta}} \right\rangle \nonumber \\
    &+ \frac{1}{16} \sum_{ijab} \sum_{klcd} \bigg[
       {t_{ij}^{ab}}^{\dagger} t_{kl}^{cd} \left\langle \Phi_{ij}^{ab} \bigg| \frac{\partial \Phi_{kl}^{cd}}{\partial H_{\beta}} \right\rangle
     + {t_{ij}^{ab}}^{\dagger} \frac{\partial t_{kl}^{cd}}{\partial H_{\beta}} \left\langle \Phi_{ij}^{ab} \bigg| \Phi_{kl}^{cd} \right\rangle
       \bigg] \bigg],
\end{align}
for which $\left[ I_{\alpha\beta}^{\lambda} \right]_{int}$ includes only terms Eqs.~\eqref{AATHF}, \eqref{AAT6} - \eqref{AAT9}.
Additionally, differentiation of the normalization factor, $N$, yields
\begin{align}
       \frac{\partial N}{\partial \chi}
    =- \frac{1}{8} \left( {1 + \frac{1}{4} \sum_{ijab} {t_{ij}^{ab}}^{\dagger} t_{ij}^{ab}} \right)^{-\frac{3}{2}}
       \left( \sum_{ijab} \frac{\partial {t_{ij}^{ab}}^\dagger}{\partial \chi} t_{ij}^{ab}
     + \sum_{ijab} {t_{ij}^{ab}}^\dagger \frac{\partial t_{ij}^{ab}}{\partial \chi} \right).
\end{align}
This derivative is zero for magnetic-field perturbations because the derivative $\hat{T}$-amplitudes are pure imaginary
quantities, and thus the final two terms in parenthesis on the right-hand side exactly cancel. The AAT with fully
normalized wave functions is then given as
\begin{align} \label{ElectronicAATNorm2}
    \left[ I_{\alpha\beta}^{\lambda} \right]&_{full}
     = N^2 \bigg[ \sum_m \sum_e U_{em}^{H_\beta} \left( U_{em}^{R_{\lambda\alpha}} + \left\langle \varphi_m^{{R_{\lambda\alpha}}} | \varphi_e^{} \right\rangle \right)
     + \frac{1}{4} \sum_{ijab} \frac{\partial t{_{ij}^{ab}}^\dagger}{\partial R_{\lambda\alpha}} \frac{\partial t_{ij}^{ab}}{\partial H_{\beta}} \nonumber \\
    &+ \frac{1}{2} \sum_{ijab} \frac{\partial {t_{ij}^{ab}}^\dagger}{\partial R_{\lambda\alpha}}
       \bigg[ \frac{1}{2} \sum_{n} U_{nn}^{H_{\beta}} t_{ij}^{ab} - \sum_{k} U_{ki}^{H_{\beta}} t_{kj}^{ab} + \sum_{c} U_{ac}^{H_{\beta}} t_{ij}^{cb} \bigg] \nonumber \\
    &+ \frac{1}{2} \sum_{ijab} \frac{\partial t_{ij}^{ab}}{\partial H_{\beta}}
       \bigg[- \sum_{k} \left( U_{ki}^{R_{\lambda\alpha}} + \left\langle \varphi_i^{R_{\lambda\alpha}} | \varphi_k^{} \right\rangle \right) {t_{kj}^{ab}}^\dagger
     + \sum_{c} \left( U_{ac}^{R_{\lambda\alpha}} + \left\langle \varphi_c^{R_{\lambda\alpha}} | \varphi_a^{} \right\rangle \right) {t_{ij}^{cb}}^\dagger \bigg] \nonumber \\
    &+ \frac{1}{2} \sum_{ijab} {t_{ij}^{ab}}^\dagger
       \bigg[ \sum_{mk} U_{km}^{H_{\beta}} \left( U_{im}^{R_{\lambda\alpha}} + \left\langle \varphi_m^{R_{\lambda\alpha}} | \varphi_i^{} \right\rangle \right) t_{kj}^{ab}
     + \sum_{ec} U_{ec}^{H_{\beta}} \left( U_{ea}^{R_{\lambda\alpha}} + \left\langle \varphi_a^{R_{\lambda\alpha}} | \varphi_e^{} \right\rangle \right) t_{ij}^{cb} \nonumber \\
    &+ \frac{1}{2} \sum_{m} \sum_{e} U_{em}^{H_{\beta}} \left( U_{em}^{R_{\lambda\alpha}} + \left\langle \varphi_m^{R_{\lambda\alpha}} | \varphi_e^{} \right\rangle \right) t_{ij}^{ab}
     - \sum_{m} \sum_{e} U_{ej}^{H_{\beta}} \left( U_{em}^{R_{\lambda\alpha}} + \left\langle \varphi_m^{R_{\lambda\alpha}} | \varphi_e^{} \right\rangle \right) t_{im}^{ab} \nonumber \\
    &- \sum_{m} \sum_{e} U_{bm}^{H_{\beta}} \left( U_{em}^{R_{\lambda\alpha}} + \left\langle \varphi_m^{R_{\lambda\alpha}} | \varphi_e^{} \right\rangle \right)  t_{ij}^{ae} \bigg] \bigg] \nonumber \\
    &+ N \frac{\partial N}{\partial R_{\lambda\alpha}} \bigg[ \sum_n U_{nn}^{H_\beta}
     + \frac{1}{2} \sum_{ijab} {t_{ij}^{ab}}^\dagger
       \bigg[ \frac{1}{2} \sum_{n} U_{nn}^{H_{\beta}} t_{ij}^{ab} - \sum_{k} U_{ki}^{H_{\beta}} t_{kj}^{ab} + \sum_{c} U_{ac}^{H_{\beta}} t_{ij}^{cb} \bigg] \nonumber \\
    &+ \frac{1}{4} \sum_{ijab} {t_{ij}^{ab}}^{\dagger} \frac{\partial t_{ij}^{ab}}{\partial H_{\beta}} \bigg].
\end{align}

\section{Computational Details}

We have implemented the analytic-gradient scheme for computing the AAT described above in the open-source Python package
apyib \cite{apyib}. To validate our implementation, we compared the fully normalized AATs with those produced by MagPy which
computes MP2 AATs using a finite-difference procedure \cite{magpy}. Both of these codes use the Psi4 quantum chemistry
package to provide the necessary integrals\cite{PSI4}.  We should note that, in order to compare finite-difference and
analytic formulations, we must assume that the perturbed HF orbitals are canonical, and thus we include $U_{pq}^\chi$ CPHF
coefficients corresponding to both dependent and independent pairs.  We will consider the use of non-canonical perturbed
MOs to simplify the computations in future work.\cite{Handy85:MP2Hessians}  

We carried out comparisons between the analytic and finite-difference AATs using several small molecular test cases,
including the hydrogen molecule dimer, water, and \hhoo\ with multiple basis sets.  However, we have chosen \hhoo\ in
conjunction with the 6-31G basis as a representative example to demonstrate the correspondence between the two approaches.
For the finite difference procedure, geometric displacements and magnetic field perturbations were set to $10^{-4}$ a.u.
Additionally, we converged SCF energies to $10^{-13}$ a.u. for both analytic- and numerical-derivative calculations.

Furthermore, we carried out computations of rotatory strengths and corresponding VCD spectra for \smo\ with the 6-31G,
6-31G(d), cc-pVDZ, and aug-cc-pVDZ basis sets\cite{Hehre1969, Ditchfield1971, Hehre1972, Hariharan1973, Dunning1989,
Pritchard2019, Feller1996} using both a common geometry / common Hessian scheme and one where the geometries
and Hessians were optimized and computed at the same level as the rotatory strengths. The CFOUR quantum chemistry program
was used for optimizing geometries as well as for computing the Hessian and APTs \cite{CFOUR}. All calculations for \smo\
were converged to $10^{-10}$ a.u.\ for the energy and $10^{-8}$ a.u.\ for the gradients.  Cartesian geometries for all data
reported in this work are available in the Supporting Information (SI).  All electrons were correlated in all calculations
reported here.

\section{Results and Discussion}

\subsection{Comparison Between Analytic and Numerical Differentiation}

The AAT obtained for \hhoo\ using the analytic-gradient method discussed above agrees closely with that obtained using the
finite-difference procedure, and can be observed from the data in Table \ref{h2o2 6-31G}.  The largest discrepancies in
individual AAT elements between the two procedures are on the order of $10^{-7}$ a.u., with most around $10^{-9}$.  This
result is not unexpected as we observe similar differences between the AATs computed at the HF level of theory using these two
approaches.  We believe this provides strong support for the correctness of both our working equations and our
implementation.

\begin{table}[h!]
    \footnotesize
        \caption{Electronic MP2 AATs (a.u.) for \hhoo\ computed with the 6-31G
        basis using analytic-gradient methods and finite-difference procedures.}
    \label{h2o2 6-31G}
    \begin{tabular}{@{}lccccccc}
    \toprule
    & \multicolumn{3}{c}{Analytic} & \hspace{1pt} & \multicolumn{3}{c}{Finite-difference} \\
    \cmidrule{2-4} \cmidrule{6-8}
    & $B_x$ & $B_y$ & $B_z$ & & $B_x$ & $B_y$ & $B_z$ \\
    \midrule
    H$_{1x}$ & 0.0060845318 &  0.0243783889 &  0.0065127385 & &  0.0060845320 &  0.0243783887 &  0.0065127390 \\
    H$_{1y}$ & 0.0385652752 & -0.1544999813 &  0.3151731127 & &  0.0385652752 & -0.1544999797 &  0.3151731091 \\
    H$_{1z}$ &-0.0693503855 & -0.1656548215 &  0.1578557162 & & -0.0693503859 & -0.1656548213 &  0.1578557139 \\
    H$_{2x}$ & 0.0060845318 &  0.0243783889 & -0.0065127385 & &  0.0060845319 &  0.0243783886 & -0.0065127383 \\
    H$_{2y}$ & 0.0385652752 & -0.1544999813 & -0.3151731127 & &  0.0385652752 & -0.1544999799 & -0.3151731086 \\
    H$_{2z}$ & 0.0693503855 &  0.1656548215 &  0.1578557162 & &  0.0693503859 &  0.1656548214 &  0.1578557132 \\
    O$_{3x}$ &-0.0136643683 &  0.1131764309 & -0.1508660802 & & -0.0136643651 &  0.1131764020 & -0.1508660412 \\
    O$_{3y}$ &-0.0449758129 & -0.0553557340 &  1.1585427903 & & -0.0449758045 & -0.0553557163 &  1.1585425174 \\
    O$_{3z}$ & 0.1014871485 & -1.0889005631 &  0.0597385980 & &  0.1014871294 & -1.0889003119 &  0.0597385786 \\
    O$_{4x}$ &-0.0136643683 &  0.1131764309 &  0.1508660802 & & -0.0136643651 &  0.1131764030 &  0.1508660409 \\
    O$_{4y}$ &-0.0449758129 & -0.0553557340 & -1.1585427903 & & -0.0449758045 & -0.0553557172 & -1.1585425196 \\
    O$_{4z}$ &-0.1014871485 &  1.0889005631 &  0.0597385980 & & -0.1014871295 &  1.0889003110 &  0.0597385803 \\
    \bottomrule
    \end{tabular}
\end{table}

As noted by Amos \textit{et al.}\cite{Amos1987}, the advantage of the analytic approach over numerical differentiation in
computing the AATs is already substantial even at the HF level of theory.  In their work, they observed that the analytic
formulation for HF AATs required only approximately two to three times more computational effort than a single SCF
calculation, whereas the numerical approach requires multiple SCF calculations --- $(3M + 3) \times 2$, where $M$ is the
number of atoms --- as well as the need for complex arithmetic.

The differences between the numerical- and analytic-derivative approaches are exacerbated at the MP2 level. For the
finite-difference procedure, evaluation of contributions involving the overlap between derivatives of doubly excited
determinants in the bra and the ket such as the last term on the right-hand side of Eq.~\eqref{MP2AATs} require
computations of the overlap between doubly excited determinants represented in different MO basis sets due to displacements
of $\delta R_{\lambda\alpha}$ for the bra wave function and of $\delta H_\beta$ for the ket wave function.  For each
combination of doubly excited determinants, this requires computation of the determinant of the overlap matrix between the two
basis sets (with rows and columns rearranged to correspond to each double excitation).  Given that there are $n_o^2 n_v^2$
doubly excited determinants each for the bra and the ket (where $n_o$ and $n_v$ are the number of occupied and virtual MOs,
respectively), and that each matrix determinant requires $n_o^3$ computational effort, the total cost of a single AAT
element at the MP2 level using numerical differentiation requires $n_o^7 n_v^4 \approx {\cal O}(N^{11})$, in addition to the
need for complex arithmetic.  

By contrast, the analytic-derivative approach requires evaluation of Eq.~\eqref{ElectronicAATNorm2}, which scales at most as
$n_o^2 n_v^4 \approx {\cal O}(N^6)$ and avoids complex wave function representations.  While our implementation is not yet
optimal, the computation of the entire MP2 AAT for \hhoo\ with the 6-31G basis set required only a few seconds using the
analytic-derivative approach and all electrons active, whereas the numerical differentiation required several hours for each
tensor element, even with the oxygen $1s$ core orbitals frozen.  The new analytic-gradient formulation makes possible MP2
VCD simulations for larger molecules and basis sets, such as those below for \smo\ which are impractical using the numerical
approach.

\subsection{MP2 VCD Spectrum of (\textit{S})-Methyloxirane}

\begin{figure}
    \centering
    \includegraphics[width=\textwidth]{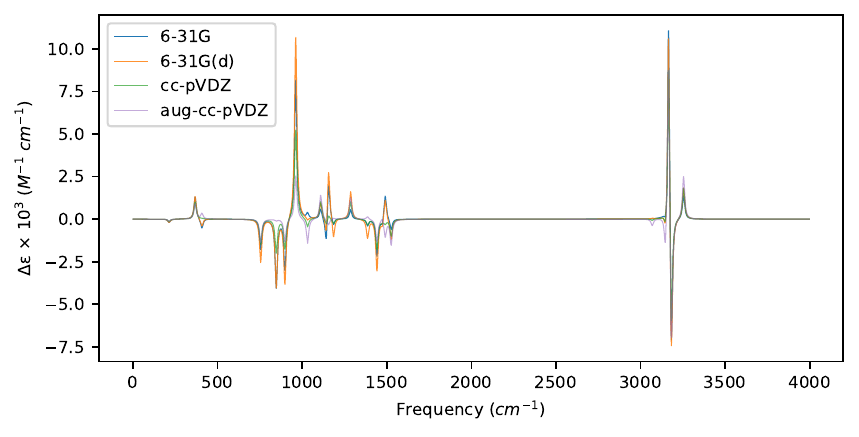}
        \caption{VCD spectra of \smo\ computed at the MP2 level of theory for several basis sets using a common optimized
            geometry and Hessian obtained at the MP2/aug-cc-pVDZ level of theory.}
    \label{SMO_Common}
\end{figure}

\begin{figure}
    \centering
    \includegraphics[width=\textwidth]{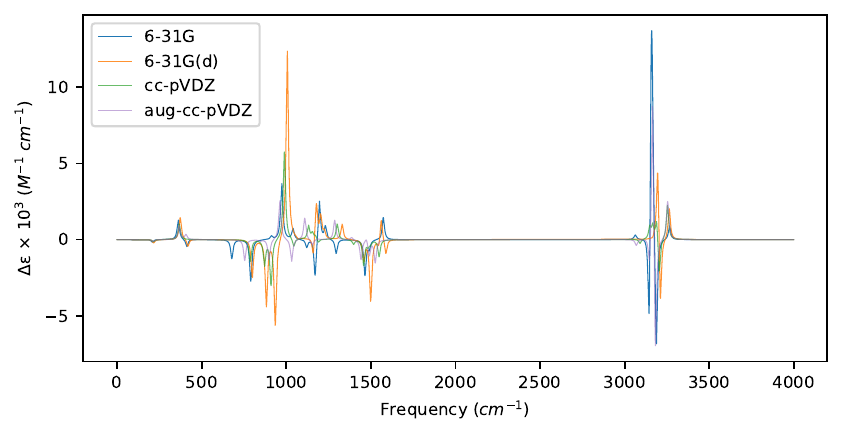}
	\caption{VCD spectra of \smo\ computed at the MP2 level of theory for several basis sets using optimized geometries
	and Hessians computed at the same level as the rotatory strengths.}
    \label{SMO_Optimized}
\end{figure}

Using this new formulation of the AATs, we have computed MP2-level VCD rotatory strengths for \smo\ for the 6-31G, 6-31G(d),
cc-pVDZ, and aug-cc-pVDZ basis sets using two different approaches.  Harmonic vibrational frequencies and rotatory strengths
are reported for \smo\ in Table \ref{methyloxirane_aug-cc-pVDZ_geom} with a common geometry and common Hessian computed
at the MP2/aug-cc-pVDZ level of theory. In addition, the corresponding VCD spectra are presented in Fig.~\ref{SMO_Common}.
Similarly, vibrational frequencies and rotatory strengths computed using geometries and Hessians computed at the same level
as the rotatory strengths are reported in Table \ref{methyloxirane_optimized_geom} with the spectra displayed in
Fig.~\ref{SMO_Optimized}. The choice of these modest basis sets helps to elucidate the variation in the MP2 rotatory
strengths upon the addition of polarization functions on carbon and oxygen, polarization functions on hydrogen, and diffuse
functions.  In addition, the choice of using a common geometry/Hessian vs.\ separate geometries/Hessians allows us to
separate structural vs.\ electronic effects in the rotatory strengths.

\begin{table}[h]
    \footnotesize
    \caption{VCD rotatory strengths of \smo\ computed at the MP2 level of theory
            for several basis set using a common optimized geometry and Hessian obtained at the MP2/aug-cc-pVDZ level.}
    \sisetup{table-format = 2.4, table-alignment-mode = format}
    \label{methyloxirane_aug-cc-pVDZ_geom}
    \begin{tabular*}{\textwidth}{@{\extracolsep{\fill}}
        %S[table-alignment-mode = marker]
        S[table-alignment-mode = none]
            c
        S[table-number-alignment = right]
        S[table-number-alignment = right]
        S[table-number-alignment = right]
        S[table-number-alignment = right]
        S[table-number-alignment = right]
        @{}}
    \toprule
{Frequency} & \hspace{1pt} & \multicolumn{4}{c}{{Rotatory Strength}} \\
{(cm$^{-1}$)} & & \multicolumn{4}{c}{{($10^{-44}$ esu$^2$ cm$^2$)}} \\
    \cmidrule{0-0} \cmidrule{3-6}
    & & {6-31G} & {6-31G(d)} & {cc-pVDZ} & {aug-cc-pVDZ} \\
    \midrule
    3254 & &  2.495 &   3.302 &   3.281 &   4.506 \\
    3181 & & -17.108 & -17.513 & -14.127 & -15.884 \\
    3164 & &  37.584 &  33.418 &  27.892 &  25.338 \\
    3160 & & -17.312 & -12.981 & -11.250 &  -7.184 \\
    3147 & &  -0.911 &  -1.928 &  -1.444 &  -4.229 \\
    3068 & &  -0.024 &   0.090 &  -0.167 &  -0.724 \\
    1526 & &  -2.477 &  -4.611 &  -3.705 &  -5.622 \\
    1490 & &   5.876 &   5.177 &  -0.835 &  -3.953 \\
    1473 & &  -1.502 &  -1.705 &  -0.408 &   1.037 \\
    1442 & &  -8.791 & -12.149 &  -7.154 &  -5.205 \\
    1386 & &  -1.543 &  -4.539 &  -1.347 &   0.702 \\
    1286 & &   2.527 &   7.301 &   4.633 &   5.736 \\
    1185 & &  -2.145 &  -6.069 &  -1.366 &   0.277 \\
    1156 & &  12.106 &  15.764 &   1.350 &  -1.706 \\
    1142 & &  -9.001 &  -7.326 &  -2.005 &  -0.683 \\
    1109 & &   3.177 &   4.890 &   5.447 &   7.494 \\
    1031 & &   1.739 &  -1.030 &  -2.964 &  -8.265 \\
     961 & &  49.773 &  64.918 &  31.811 &  15.573 \\
     898 & & -19.628 & -25.045 & -11.611 &  -8.674 \\
     846 & & -27.450 & -27.241 & -13.475 &  -0.573 \\
     754 & & -13.513 & -19.358 & -11.231 & -10.496 \\
     406 & &  -8.237 &  -6.523 &   0.230 &   4.505 \\
     367 & &  21.043 &  21.287 &  15.800 &  14.207 \\
     212 & &  -5.346 &  -5.160 &  -3.644 &  -3.375 \\
    \bottomrule
    \end{tabular*}
\end{table}

\begin{table}[h]
    \footnotesize
        \caption{VCD rotatory strengths of \smo\ computed at the MP2 level of theory
            for several basis sets using optimized geometries and Hessians computed at the same level as the rotatory
                strengths.  The units of frequency are cm$^{-1}$ and rotatory strengths are $10^{-44}$ esu$^2$ cm$^2$.}
    \label{methyloxirane_optimized_geom}
    \sisetup{table-format = 2.4, table-alignment-mode = format}
    \begin{tabular*}{\textwidth}{@{\extracolsep{\fill}}
        S[table-alignment-mode = none]
        S[table-number-alignment = right]
            c
        S[table-alignment-mode = none]
        S[table-number-alignment = right]
            c
        S[table-alignment-mode = none]
        S[table-number-alignment = right]
            c
        S[table-alignment-mode = none]
        S[table-number-alignment = right]
        @{}}
    \toprule
    \multicolumn{2}{c}{{6-31G}} & \hspace{1pt} & \multicolumn{2}{c}{{6-31G(d)}} & \hspace{1pt} &
    \multicolumn{2}{c}{{cc-pVDZ}} & \hspace{1pt} & \multicolumn{2}{c}{{aug-cc-pVDZ}} \\
    \cmidrule{0-1} \cmidrule{4-5} \cmidrule{7-8} \cmidrule{10-11}
{Freq.} & {Rot. Str.} & & {Freq.} & {Rot. Str.} & & {Freq.} & {Rot. Str.} & & {Freq.} & {Rot. Str.} \\
    \midrule
     3265 &   1.636 & & 3263 &   3.672 & & 3253 &   4.073 & & 3254 &   4.506 \\
     3187 & -14.649 & & 3210 &  -9.250 & & 3204 &  -4.480 & & 3181 & -15.884 \\
     3163 &  15.085 & & 3196 &   9.875 & & 3189 &   2.766 & & 3164 &  25.338 \\
     3156 &  19.904 & & 3180 &   0.509 & & 3165 &   1.637 & & 3160 &  -7.184 \\
     3145 & -17.195 & & 3167 &  -0.217 & & 3148 &   1.437 & & 3147 &  -4.229 \\
     3063 &   0.557 & & 3108 &  -0.010 & & 3091 &  -0.508 & & 3068 &  -0.724 \\
     1573 &   6.154 & & 1588 &  -3.549 & & 1548 &  -4.130 & & 1526 &  -5.622 \\
     1573 &  -0.692 & & 1561 &   5.633 & & 1498 &  -0.985 & & 1490 &  -3.953 \\
     1557 &  -0.203 & & 1546 &  -2.039 & & 1483 &  -1.420 & & 1473 &   1.037 \\
     1490 &  -2.111 & & 1498 & -15.452 & & 1457 &  -6.593 & & 1442 &  -5.205 \\
     1465 &  -9.245 & & 1458 &  -5.317 & & 1397 &  -1.216 & & 1386 &   0.702 \\
     1294 &  -4.141 & & 1330 &   4.396 & & 1301 &   4.587 & & 1286 &   5.736 \\
     1231 &   3.987 & & 1225 &   2.176 & & 1190 &  -1.165 & & 1185 &   0.277 \\
     1195 &  13.263 & & 1203 &   7.003 & & 1172 &   1.077 & & 1156 &  -1.706 \\
     1170 & -12.766 & & 1178 &  11.823 & & 1154 &   2.005 & & 1142 &  -0.683 \\
     1120 &  -2.652 & & 1159 &  -6.617 & & 1131 &   4.624 & & 1109 &   7.494 \\
     1039 &   3.918 & & 1070 &  -0.714 & & 1044 &  -3.006 & & 1031 &  -8.265 \\
      973 &  22.078 & & 1005 &  71.797 & &  988 &  33.915 & &  961 &  15.573 \\
      912 &   1.363 & &  934 & -35.039 & &  910 & -19.030 & &  898 &  -8.674 \\
      790 & -19.944 & &  882 & -28.207 & &  871 & -11.066 & &  846 &  -0.573 \\
      678 & -10.592 & &  799 & -17.916 & &  787 & -10.455 & &  754 & -10.496 \\
      412 &  -6.849 & &  418 &  -6.702 & &  409 &  -0.073 & &  406 &   4.505 \\
      361 &  21.032 & &  372 &  22.692 & &  367 &  16.270 & &  367 &  14.207 \\
      208 &  -4.897 & &  216 &  -5.916 & &  214 &  -3.874 & &  212 &  -3.375 \\
    \bottomrule
    \end{tabular*}
\end{table}

We observe in Table \ref{methyloxirane_aug-cc-pVDZ_geom} that, even using a common geometry and Hessian, there is
considerable basis-set dependence in the rotatory strengths, including multiple instances of sign changes relative to the
aug-cc-pVDZ basis set results.  For example, for the normal modes at 1156 cm$^{-1}$ (hydrogen rocking vibration), 1185
cm$^{-1}$ (also a hydrogen rocking vibration), and 1386 cm$^{-1}$ (an in-phase bending vibration primarily associated with
the hydrogens on the methyl group), the three smallest basis sets yield the opposite sign rotatory strengths from the
aug-cc-pVDZ basis set.  These same sign differences, though weak, can be seen in the VCD spectrum in Fig.~\ref{SMO_Common}.
For the stronger rotatory strengths, we observe no sign changes across the basis sets, though the significance of diffuse
functions can still be seen in the mode at 846 cm$^{-1}$ (a ring breathing vibration), where the aug-cc-pVDZ basis set
yields a rotatory strength factor of 24-48 times smaller than the other sets.  Interestingly, the five highest frequency
modes (all C$-$H stretching motions) are relatively well described by all the basis sets considered here.

When we use geometries and Hessians computed at the same level of theory as the rotatory strengths, as shown in Table
\ref{methyloxirane_optimized_geom} and Fig.~\ref{SMO_Optimized}, we observe much greater variation in the resulting spectra,
even in the high-frequency regime.  For example, the C$-$H stretching vibrations at 3160 and 3147 cm$^{-1}$ with the
aug-cc-pVDZ basis set exhibit a sign change compared to the cc-pVDZ basis set (and compared to all three smaller basis sets
for the former).  Indeed, we observe sign reversals between cc-pVDZ and aug-cc-pVDZ for eight vibrational modes, though
primarily for relatively weak rotatory strengths.  Even for modes for which the signs are consistent, however, we observe
large changes in the magnitude of the rotatory strength among the basis sets.  For example, for the five modes with an
absolute rotatory strength greater than 10.0$\times$$10^{-44}$ esu$^2$ cm$^2$ at the MP2/aug-cc-pVDZ level, three exhibit an
intensity shift larger than a factor of two between MP2/cc-pVDZ and MP2/aug-cc-pVDZ.

\section{Conclusion} In this work, we have derived and implemented an analytic-gradient method for computing VCD AATs at the
MP2 level of theory. We have compared this method with that of our recently developed numerical-differentiation procedure
--- which is several orders of magnitude more computationally expensive and less precise than our new formulation --- and
obtained very good agreement between the two approaches. Using this new implementation, we also report the first MP2 level
VCD spectra of \smo\ using several modest basis sets.\cite{Shumberger2023}.

This work is expected to allow the simulation of fully analytic MP2-level VCD spectra for a wide array of molecules and
basis sets, and it opens the door to future implementations of VCD at even higher levels of theory. Although the present
effort does not include the use of GIAOs, resulting in origin-dependent rotatory strengths, our future work will be
directed towards the development of a gauge- and origin-invariant formulation.

\clearpage

\section{Supporting Information} \label{si}

Atomic coordinates of the test molecules are provided.

\section{Acknowledgements} \label{ack}

TDC was supported by the U.S.\ National Science Foundation via grant CHE-2154753 and BMS by grant DMR-1933525.
The authors are grateful to Advanced Research Computing at Virginia Tech for providing computational resources
that have contributed to the results reported within the paper.

\bibliography{references}

\newpage

\end{document}